# Volume and Surface-Enhanced Volume Negative Ion Sources

*M.P. Stockli*

Spallation Neutron Source, Oak Ridge National Laboratory, Oak Ridge, TN 37830, USA

**Abstract**

$H^-$ volume sources and, especially, caesiated $H^-$ volume sources are important ion sources for generating high-intensity proton beams, which then in turn generate large quantities of other particles. This chapter discusses the physics and technology of the volume production and the caesium-enhanced (surface) production of $H^-$ ions. Starting with Bacal's discovery of the $H^-$ volume production, the chapter briefly recounts the development of some $H^-$ sources, which capitalized on this process to significantly increase the production of $H^-$ beams. Another significant increase was achieved in the 1990s by adding caesiated surfaces to supplement the volume-produced ions with surface-produced ions, as illustrated with other $H^-$ sources. Finally, the focus turns to some of the experience gained when such a source was successfully ramped up in $H^-$ output and in duty factor to support the generation of 1 MW proton beams for the Spallation Neutron Source.

## 1 Introduction

This chapter discusses the production of large quantities of negative hydrogen ions ($H^-$) to form small-diameter $H^-$ beams, which then are accelerated to high energies to create powerful proton beams. Colliding with other beams or impacting on specially engineered targets, these proton beams will then produce high yields of the desired secondary particles, such as neutrons in the 1 MW Spallation Neutron Source (SNS) [1]. The discussion focuses on the successful high-yield $H^-$ production with volume and caesiated-volume sources, such as the SNS $H^-$ source [2], the DESY $H^-$ source at the Deutsches Electronen Synchrotron [3], the J-PARC $H^-$ source at the Japan Proton Accelerator Research Complex [4], and a few others that were important for those developments. It excludes the compact surface plasma sources, which are the topic of a chapter by D. Faircloth.

The SNS $H^-$ source was specified with requirements that exceeded the capabilities of all existing $H^-$ sources [5]. Although there were $H^-$ sources successfully producing more than the required 50 mA $H^-$ beam current, those sources were operated at duty factors that were orders of magnitude smaller. Although there were $H^-$ sources successfully operating with duty factors exceeding the 6–7% required for SNS, those sources were producing only a fraction of the required 50 mA. At that time, it was commonly assumed that increasing the duty factor would inversely decrease the lifetime. Accordingly, the specified 6–7% duty factor made the requirement for three-week-long, maintenance-free service cycles [6] a leap of faith.

In 1994 Lawrence Berkeley National Laboratory (LBNL) hosted a workshop to evaluate options and issues for an $H^-$ source, a low energy beam transport (LEBT) system, and beam chopping for SNS [7]. Although a broad range of required R&D was identified, the radio-frequency (RF)-driven, Cs-enhanced, multicusp $H^-$ source, which was developed at LBNL and tested at SSC, looked very promising [8]. The Cs cartridges, containing only a few milligrams of Cs, were very successful in tests at LBNL, and later at the Superconducting Super Collider (SSC) [9], although the duty factor was 60 times smaller than the SNS requirement. Also, to intercept the co-extracted electrons before they gain prohibitively high energies, a strong dipole magnet was integrated into the outlet aperture [10], an elegant and compact, but untested, solution that remains to be fully evaluated.

Since 2009 SNS has been operating near 1 MW, with the H$^-$ ions supplied with slightly modified LBNL H$^-$ sources. These deliver daily ~230 C of H$^-$ ions, which is unprecedented for pulsed H$^-$ sources for accelerators [2].

In six-week maintenance-free service cycles the sources deliver ~10 kC or ~2.7 A h of H$^-$ ions, which is also unprecedented for pulsed H$^-$ sources for accelerators [2]. Ramping up the duty factor 25-fold over a three-year period revealed many unprecedented issues, which had to be understood and mitigated to continue the ramp-up with an acceptable availability.

Accordingly, the CERN Accelerator School (CAS) programme committee requested the inclusion of some of the SNS lessons learned, including the $Cs_2CrO_4$ system, and scaling rules for H$^-$ sources.

## 2    Historical background

Negative ion sources for accelerators were initially developed to allow the sources to be placed on easily accessible, service- and control-friendly, isolated platforms that are negatively charged to a few tens of kilovolts. So called 'tandems' accelerate the negative ions to a highly charged positive high-voltage (HV) terminal, strip at least two electrons, and then accelerate the resulting positive ions back to ground, which doubles the energy of singly charged, or multiplies the energy of multiply charged, heavy ions.

Double-stripping H$^-$ ions produces protons, which travel with opposite curvature in magnetic fields. This change of direction is used in cyclotrons to extract protons without the need of an extraction channel, which is prone to arcing and strong activation.

Accumulator rings can easily accumulate identical particles as long as the particle bunches remain separated in space so that kickers can be activated between bunches. This limit can be overcome by stripping negative ions inside an injection magnet, after which the now positive ions join the trajectories of the already stored positive ions. For example, SNS accelerates ~1000 beamlets of ~40 mA H$^-$ ions, which are (mostly) double stripped in the injection magnet. Here they join the proton beam in the accumulator ring, which growths to several tens of amperes as a result. This allows the delivery of all protons to the Hg spallation target in less than 1 μs, although it takes ~1 ms to produce those protons.

The production of negative ions was a specialty limited to a few accelerator laboratories. However, that changed when it became clear that magnetically confined fusion needed neutral beam heating to reach the temperatures required for fusion reactors. Many researchers and laboratories started programmes to model, study, or develop large negative ion sources and the required neutralizers for a new class of high-voltage, high-current accelerators.

The extensive research on H$^-$ production and the development of H$^-$ sources are primarily documented in the proceedings of the 'International Symposia on the Production and Neutralization of Negative Ions and Beams' (PNNIB), which were mostly held at Brookhaven National Laboratory. Except for the first two, all proceedings are available as conference proceedings from the American Institute of Physics (AIP). Reduced funding in the 1990s reduced the interest and the symposia stopped in 1997. The symposium was restarted in 2002 in part to support the unprecedented challenge SNS was facing. In 2008 the biennial symposium was renamed as the 'International Symposium on Negative Ions, Beams and Sources' (NIBS), while the proceedings continue to be published by AIP.

## 3  The volume production of H⁻ ions

The second electron on an H⁻ ion is bound with 0.75 eV, about 20 times less than the ~15 eV ionization energy of hydrogen atoms or molecules. This mismatch makes the direct formation of H⁻ ions in plasma rare events, such as the radiative electron capture by hydrogen atoms, which peaks with ~$10^{-18}$ cm² near 1 eV [11]. Even less likely is the dissociative attachment of fast electrons (>7 eV) to ground-state hydrogen molecules, yielding H⁻ ions with a cross-section of ~$10^{-20}$ cm² [12].

Accordingly, it was rather surprising when in 1977 M. Bacal found signals of large negative ion populations in hydrogen plasma. Years of research have shown that the dominant production occurs when highly rovibrationally excited hydrogen molecules ($4 \leq v \leq 9$) disintegrate after colliding with slow electrons (~1 eV), having cross-sections of up to ~$10^{-15}$ cm² [11, 12].

Highly rovibrationally excited molecules are easily produced with fast electrons (>20 eV), with cross-sections of up to $5 \times 10^{-18}$ cm² [12, 13]. Unfortunately, such fast electrons (>5 eV) destroy H⁻ ions rather rapidly, with cross-sections of up to $4 \times 10^{-15}$ cm² [11, 12], severely limiting the H⁻ population in hot plasma. H⁻ ions have much longer lifetimes in cold plasma because of their ionization threshold at 0.75 eV.

This problem of production and destruction was overcome with tandem sources [14, 15], which contain a magnetic filter between the plasma-producing region and the ion outlet, as shown in Fig. 1. The magnetic dipole field generates the Lorentz force, which is perpendicular and proportional to the velocity of charged particles. This force turns fast electrons around, returning them towards the filament. The Lorentz force has fewer effects on slow, collisional electrons and even slower protons, which diffuse through the filter field and generate much colder plasma near the ion outlet, so extending the lifetime of the locally generated H⁻ ions.

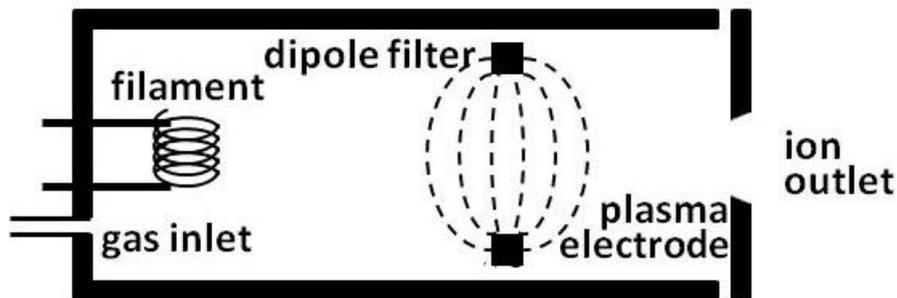

**Fig. 1**: Schematic of a tandem source with a magnetic filter

Filter magnets keep the destructive hot electrons away from the extraction region. However, the loosely bound second electron can also be detached by photons (~$10^{-17}$ cm²) [12] or in collisions with H atoms or molecules (~$10^{-15}$ cm²) [11, 12]. The dominant loss is probably the mutual neutralization with protons, especially with slow protons (e.g., ~$7 \times 10^{-14}$ cm² for 0.5 eV H⁺) [12]. Accordingly, H⁻ ions formed near the source outlet have a much better chance to be extracted, and therefore the filter fields are now located close to the outlet.

## 4  Volume negative ion sources

Many so-called 'volume' negative ion sources have been developed that deliver predominantly H⁻ ions produced in the volume near the outlet. Foremost is 'Camembert' [16] at the Ecole Polytechnique in Palaiseau, France, which is extensively used to study the volume production of H⁻ and D⁻ ions [17]. A filament discharge generates plasma, which is confined by multicusp magnets and features a plasma electrode [14].

In the 1980s LBNL started to develop filament-driven H⁻ sources. In the late 1980s a small source with a 75 mm inner diameter was developed using all permanent magnets as shown in Fig. 2(a) [18]. Fourteen rows of water-cooled $SmCo_5$ magnets provide radial confinement, and another four rows in the back flange provide rear confinement. The filter field was generated with a pair of water-cooled permanent magnets mounted on the outlet flange. Placing a pair of $SmCo_5$ magnets 40 mm apart produced the highest H⁻ current when ~1 V was applied the plasma electrode (PE), as the outlet electrode is normally called. Up to 2 mA H⁻ peak current was extracted through a 1 mm diameter outlet when 80 kW of discharge power was applied to generate the 1 ms long pulses [18].

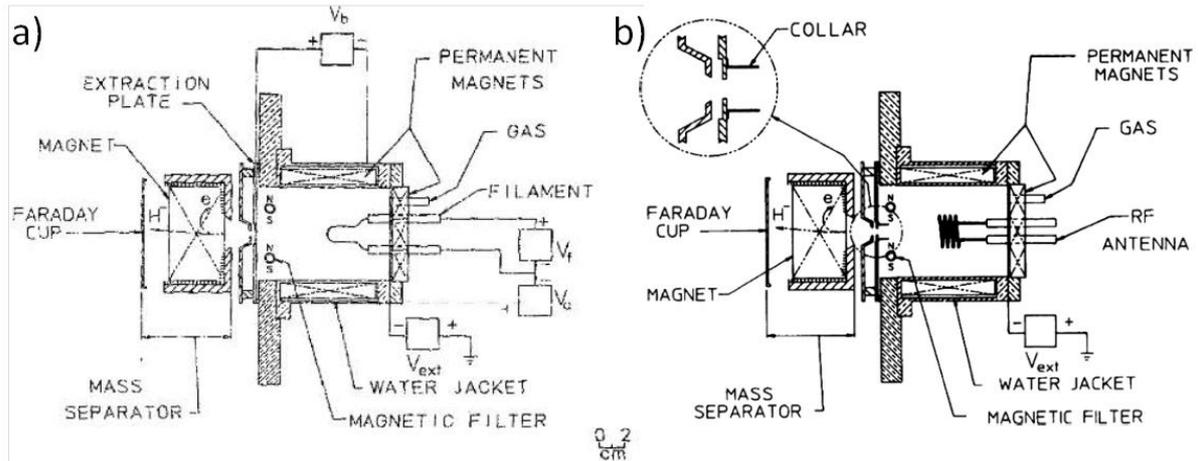

**Fig. 2**: Schematics of the small (a) filament-driven and (b) RF-driven volume H− source developed at LBNL

In 1990 LBNL started to develop an H⁻ source for the SSC. Replacing the filament in their small multicusp source by a three-turn inductive coil, shown in Fig. 2(b), increased the H⁻ current from 4.2 mA to 6 mA [19]. The outlet had 2 mm diameter and the plasma was driven with 25 kW discharges. A collar surrounded the outlet, which was shown to reduce the number of co-extracted electrons.

In 1992, the LBNL-built SSC H⁻ source met its requirement by producing 30 mA with a two-turn antenna driven by 35–45 kW RF and ignited with a W filament [20]. Studying the emittance of the H⁻ beam, it was shown that that a ~45% downstream taper of the 6.3 mm outlet yield a favourable emittance and divergence [21].

After acquiring a copy of the LBNL RF-driven H⁻ source in 1994, DESY found the antenna lifetime unacceptable [22]. Despite testing other configurations, DESY was unable to produce robust antennas, which would yield adequate H⁻ beam current with an adequate lifetime [23]. Finally, in the late 1990s DESY decided to develop an external antenna H⁻ source where the antenna is wound around an $Al_2O_3$ cylinder, which forms the wall of the plasma chamber shown in Fig. 3(a) (adapted from [23]). In addition, it was decided to reduce operational risks [24] by not using Cs.

The resulting source was very successful, producing a 40 mA H⁻ beam for 0.15 ms at 6 Hz for over 100 weeks [25]. A pulse length of 3 ms [26] was demonstrated and 60 mA peak currents were also achieved [25]. Figure 3(b) (from [25]) shows the segmented collar that was studied to elucidate the function of the collar. Briefly, the collar allows the electron density to be lowered, especially at the entrance of the collar. This draws positive ions to the collar wall, where they neutralize to form hyper-thermal neutrals, which in turn can form additional rovibrationally exited molecules, enhancing the production of H⁻ ions [25].

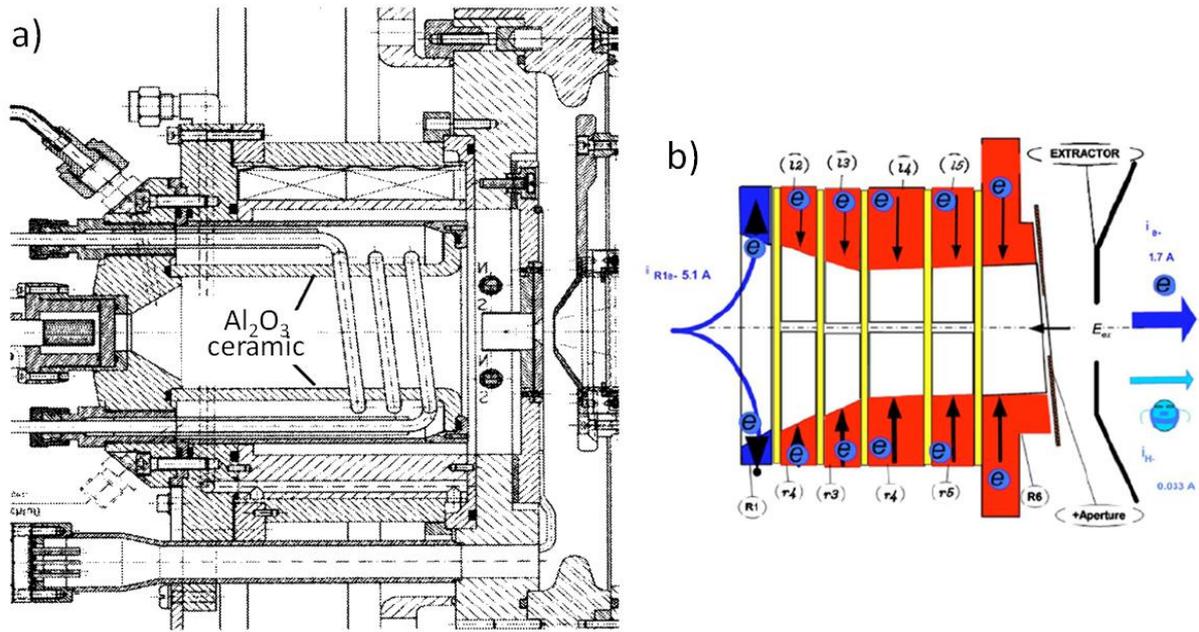

**Fig.3**: (a) Schematic and (b) detailed outlet collar of the external antenna H− source developed at DESY ((a) adapted from [23]; (b) from [25]).

In the late 1980s TRIUMF started to develop a 10 cm inner diameter, filament-driven H⁻ source to inject a continuous H⁻ beam into their cyclotron [27]. Figure 4(a) shows a cross-section through the source with two of the ten rows of $SmCo_5$ magnets that produce the plasma-confining multicusp field. At the outlet end, two diametrically opposed magnets are reversed to produce the filter field [27]. This causes the electron temperature to drop from ~2 eV in the centre of the source to <0.3 eV at the 13 mm diameter outlet, which is shown in detail in Fig. 4(b) [28, 29]. The source yields up to 15 mA of H⁻ at 20–30 kV, producing ~3 mA per kilowatt of required power [26], which is very efficient for H⁻ sources. The design has been licensed and the source is commercially available [30]. Operation near the peak current requires the filament to be replaced about every two weeks [30].

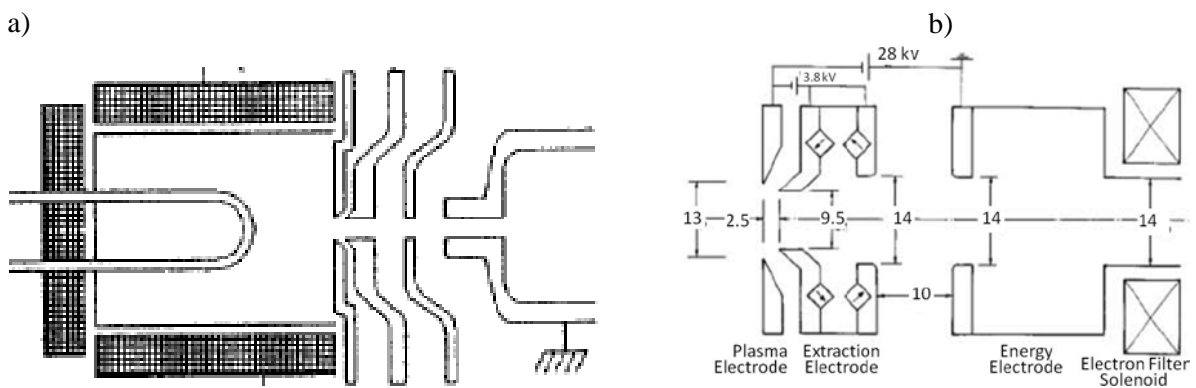

**Fig. 4**: (a) Schematic and (b) detailed extraction region of the 15 mA continuous H⁻ source developed by TRIUMF ((a) adapted from [27]; (b) adapted from [26]).

To optimize the extraction of the 'volume'-produced H⁻ ions, production 'volume' H⁻ sources have large outlets of the order of 1 cm diameter, in contrast to the small source outlets featured by the compact surface plasma sources discussed by D. Faircloth. The large outlet causes significant leakage of neutral hydrogen gas, typically 20–40 sccm (standard cubic centimetres per minute), which needs to be differentially pumped to avoid large stripping losses in the LEBT (~$10^{-15}$ cm$^2$ for $E_{H^-} > 100$ eV) [12].

## 5     The surface production of H$^-$ ions

Compared to the ~15 eV ionization energy of hydrogen atoms and molecules, it is easier to remove electrons from metal surfaces, which have work functions between 4 and 5.6 eV, as the electron affinities of surfaces are called [12]. This is relevant because the walls of the plasma container are continuously exposed to a flux of thermal hydrogen molecules, which tend to stick for a while and later thermally desorb. In addition the plasma generates significant fluxes of hyper-thermal atoms and molecules, which also impact on the surfaces. There are also fast positive hydrogen ions (H$^+$, H$_2^+$ and H$_3^+$), which are accelerated in the sheath by the plasma potential, and then Auger-neutralize when they reach the surface. Owing to their kinetic energy, the hyper-thermal neutrals and neutralized ions are likely to bounce back or alternatively sputter an adsorbed hydrogen atom or molecule with a hyper-thermal velocity.

When a hydrogen atom leaves the surface, a conduction electron near the surface can get trapped in the field of the atom, forming an H$^-$ ion. However, the surface work function is always more attractive than the 0.75 eV electron affinity of H atoms. Accordingly the dominant fraction of electrons return to the surface, especially for atoms leaving the surface slowly, thus giving the electrons time to choose.

Rasser gives a low-velocity approximation for the probability $\beta^-$ for an atom to form a negative ion:

$$\beta^-(v_\perp) \approx (2/\pi) \exp[-\pi(\Phi - S)/(2av_\perp)], \qquad (1)$$

where $v_\perp$ is the emitted atom's velocity normal to the surface, $\Phi$ is the work function of the surface, $S$ is the electron affinity of the atom, and $a$ is a decay constant [31]. The approximation shows that the probability depends exponentially on the inverse of the normal velocity, yielding very small probabilities for small velocities. Furthermore, it depends exponentially on the difference between the work function and the electron affinity, which for hydrogen is dominated by the larger work function.

Rasser's full model yields a probability that saturates for large velocities (>100 eV H) at ~4% for H on clean W(110). However, the probability drops to ~2% for ~10 eV H atoms [31]. The work function of W(110) is 5.25 eV, slightly higher than the 4.95 eV for Mo(110) [32], and accordingly Mo should yield similar percentages.

It is possible to enhance the probabilities by lowering the work function through the adsorption of alkali atoms [32]. Figure 5 shows the approximate work function of a Mo surface as a function of coverage by adsorbed Cs atoms. The coverage is measured in units of a monolayer, a one-atom thick layer of densely packed Cs atoms. The work function starts near 4.6 eV, the value of a clean, polycrystalline Mo substrate, free of Cs. As Cs is gradually added, the work function decreases and reaches a minimum of ~1.6 eV near 0.6 monolayer [32]. Adding more Cs increases the work function until an equilibrium value is reached close to 2.1 eV, the work function of bulk Cs.

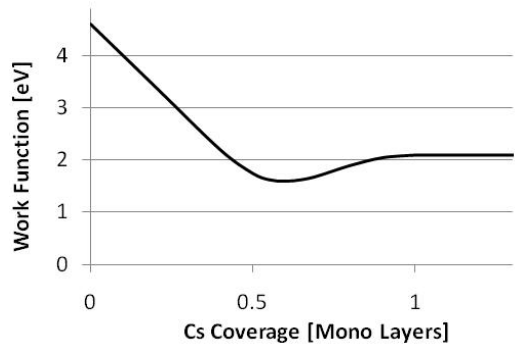

**Fig. 5:** Work function for Mo surfaces as a function of adsorbed Cs

For a W(110) surface covered with 0.5 monolayer of Cs, Rasser calculated ~40% probability for fast H atoms to form negative ions, and ~30% for 10 eV H atoms [31]. Rasser's calculations have been confirmed by measurements of up to 34% $H^-$ ions when a proton beam was reflected at small grazing angle from a caesiated W(110) surface with a 1.45 eV work function [33].

While these are impressive fractions, they apply only to $H^+$ ions that bounce back elastically after hitting a Cs atom or a heavy atom of the substrate. Molecular ions have smaller velocities, and many of the ions lose a significant fraction of their energy when scattering with the light hydrogen atoms and molecules adsorbed on the surface, some of which are sputtered.

Much smaller percentages can be expected for atoms ejected due to impacting hyper-thermal atoms and molecules. Okuyama scattered 0.11–0.14 eV hydrogen atoms from a Cs-covered Mo substrate, and found ~0.02% $H^-$ production probabilities, which are in reasonable agreement with Eq. (1) [34]. Model calculations find large fluxes of hyper-thermal atoms, and their contributions to $H^-$ production can outnumber the contributions from the ions [35].

It is important to understand that many processes contribute to the extracted $H^-$ beam in many interconnected ways. Optimizing the performance of an $H^-$ source must always be judged from the carefully measured, extracted $H^-$ beam and its emittance.

## 6     Filament-driven surface-enhanced volume $H^-$ ion sources

In the 1990s the Japan Atomic Energy Research Institute (JAERI) started to develop $H^-$ ion sources for a multipurpose high-energy proton accelerator project. The development produced a 150 mm inner diameter, multicusp $H^-$ source driven by two filaments, featuring a Mo plasma electrode with an 8 mm outlet. Adding Cs increased the extracted $H^-$ threefold, yielding up to 72 mA with 56 kW discharge power [36]. Unfortunately the lifetime was limited by the filament, which fell short of the three-week project requirement.

In the 2000s the Japan Atomic Energy Agency (JAEA) and the High Energy Accelerator Research Organization (KEK) constructed J-PARC jointly. Concerns that Cs may compromise the performance of the radio-frequency quadrupole (RFQ) accelerator drove the development of a new $H^-$ source that could operate without Cs. After developing long-lived, highly emissive $LaB_6$ filaments, optimizing the geometry and the local fields, shown in Fig. 6(a), and optimizing the Mo plasma electrode with a 9 mm outlet and the extraction system, shown in Fig. 6(b), up to 38 mA current was obtained for ~50 kW discharge power [37]. The 2.5 eV work function of $LaB_6$, the careful shaping of the plasma electrode, its coating of mostly B and some La after use [38], as well as its high operational temperature suggest that a significant contribution of surface-produced $H^-$ ions [37] can be obtained without using Cs. This source has served J-PARC with high availability, easily meeting the 1200 h lifetime requirement for the 17 mA, 0.5 ms long pulses at 25 Hz [4].

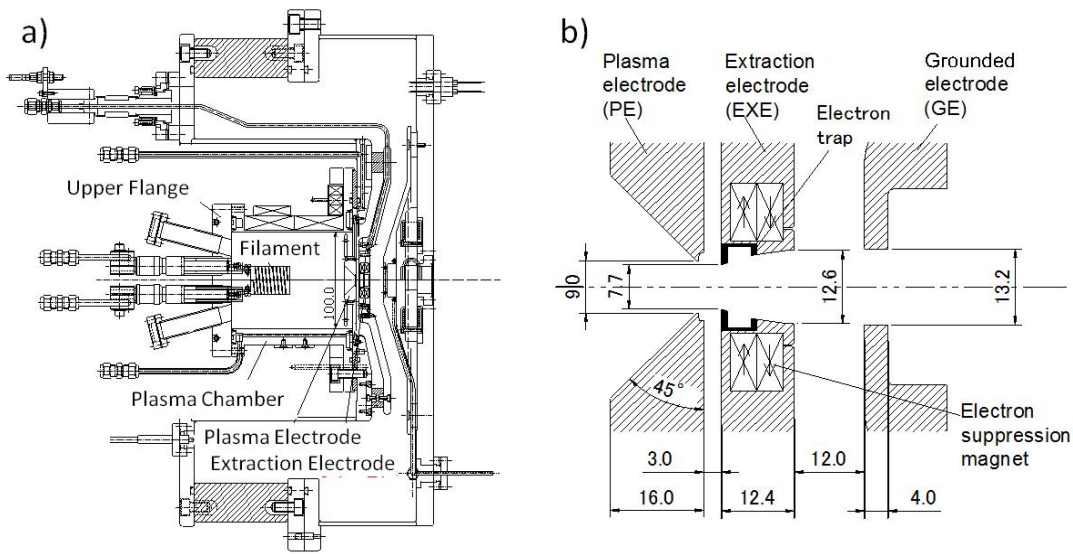

**Fig. 6**: (a) Schematic and (b) detailed extraction system of the H⁻ source developed and used at J-PARC [39]

However, 60 mA will be required to upgrade J-PARC to 1 MW with 0.5 ms pulses at 25 Hz and a three-week lifetime. Renewed filament optimizations yielded up to 45 mA [38], missing the 60 mA goal. Adding Cs yielded only a modest enhancement for the $LaB_6$ filaments [37]. Using a W filament yielded 18 mA before and 72 mA after caesiation, with a discharge power of only 16 kW. However, this level was maintained only for a limited time and there was consumption of the Cs [37].

Arc discharges sputter the filaments, which limits their lifetime. Some of the sputtered material coats the surfaces near the outlet, which in some cases can enhance the surface production of H⁻ ions. However, it can also sputter the Cs or cover the Cs layer, and therefore filament-driven H⁻ sources require a steady supply of Cs. The J-PARC experiments suggest that using Cs in filament-driven sources can only yield long lifetimes at low duty factors.

## 7  RF-driven surface-enhanced volume H⁻ ion sources

In 1992 the electron-suppressing collar of the RF-driven LBNL H⁻ source was replaced with a collar holding two Cs cartridges, shown in Fig. 7(a). This increased the H⁻ current output by a factor of three [40], converting this source to a surface-enhanced volume H⁻ source.

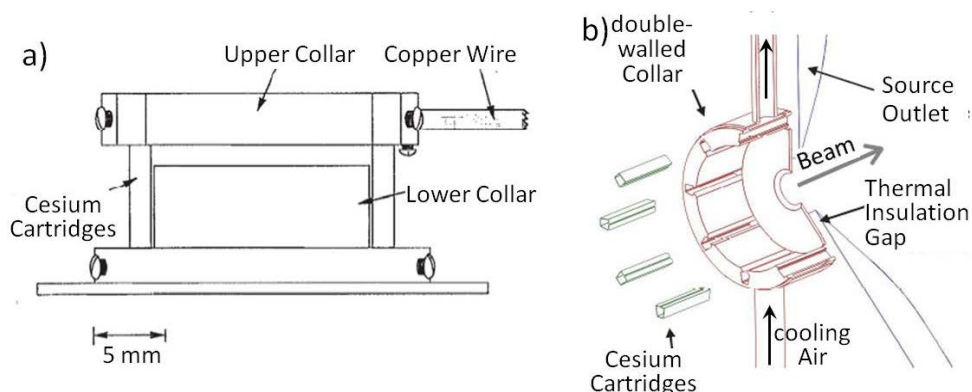

**Fig. 7:** Schematics of the (a) prototype and (b) SNS Cs collar developed at LBNL

In 1995, the same collar was tested in the SSC H⁻ source, which produced up to 100 mA and continued for three days to deliver over 80 mA for 0.1 ms at 10 Hz [9]. Again, a volume source was converted to a surface-enhanced volume H⁻ source.

In the late 1990s, LBNL started the detailed design of the SNS source, building on previous success and incorporating the advances achieved with the SSC source. Several changes had to be made to address the drastically higher duty factor of the SNS source. While the use of Cs cartridges was adopted from the SSC source, they were integrated into an air-cooled collar to control their temperature, as shown in Fig. 7(b). Increasing the collar temperature to 400°C increased the H⁻ current extracted through the 7 mm outlet from 18 mA to 56 mA for a 30 kW RF discharge, while drastically reducing the co-extracted electron current [41].

In addition, a dipole magnet was integrated into the outlet flange to drive the co-extracted electrons sideways onto the e-dump electrode, shown in Fig. 8(a) [10]. Several ignition schemes were tested, but these were abandoned in favour of a continuous 13 MHz discharge [42]. To limit the emittance growth, a very compact LEBT was designed. It comprised two electrostatic einzel lenses, with the second lens split into four quadrants to steer and chop the beam [43]. The source and LEBT were successfully commissioned at LBNL, producing 50 mA beam pulses [44]. In 2002 it was shipped to, installed and recommissioned at Oak Ridge National Laboratory (ORNL). After initial problems with high duty factors, the commissioning of the SNS accelerator was performed with short pulses at low repetition rates, which yielded ~99% availability for the ion source and LEBT system [45].

Challenges arose in 2007 after stepping up to 15 Hz and trying to extend the pulse length to 0.3 ms. However, these problems were analysed and mitigated, normally by the start of the next ~20 week run [45, 46], yielding availabilities mostly between 95 and 99%.

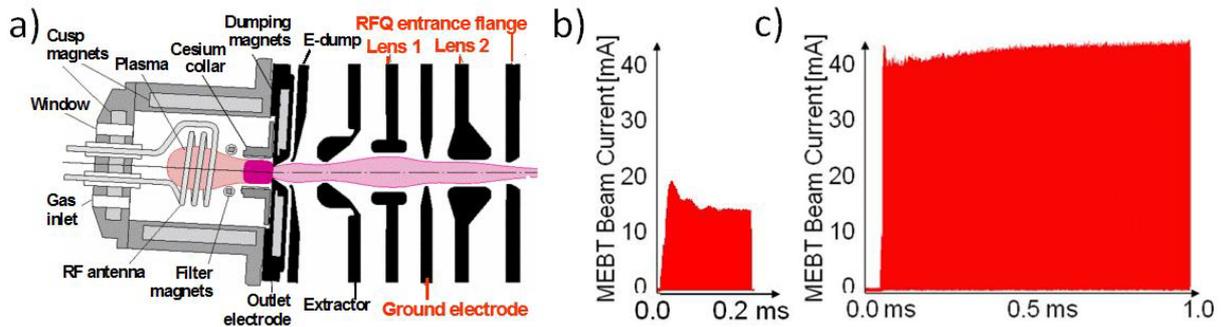

**Fig. 8:** (a) Schematic of the SNS H⁻ source and LEBT; (b) RFQ output in 2007; and (c) RFQ output in 2009

Ramping up the duty factor from ~0.5% to ~5% over a two-year period required the modification of configurations and procedures [46]. The result was an increase in the H⁻ beam current from ~13 mA in summer of 2007, as shown in Fig. 8(b), to ~40 mA in autumn of 2009, as shown in Fig. 8(c), which exceeded the 38 mA requirement for ~1 MW of beam power.

A series of interesting lessons was learned regarding the radio frequency: ~300 W of 13 MHz sustains continuous plasma, which eases [42] but does not guarantee the breakdown of the pulsed high-power plasma. The high-power plasma pulses are generated by superimposing high power at 2 MHz for the desired pulse length and repetition rate.

In 2007, when the high-power RF pulse length was extended beyond 0.1 ms, the beam current initially dropped off, as seen in Fig. 8(b). However, increasing the matching capacity increased the latter part of the current while lowering the initial overshoot, just as desired.

Ramping up the H⁻ beam current required increasing the RF power and the match accordingly, which frequently led to plasma outages. Trying to match the high-power plasma, which includes plasma inductance, increasingly mismatched the lower-inductance initial state until the induced electric fields were too weak to break down the plasma [47]. Using a compromise tune and increasing the matching network inductance reduced the problem, but did not eliminate it.

In 2009 an increasing number of plasma outages were noted towards the end of the four-week source service cycles, especially towards the end of the approximately 20-week run [47]. This was caused by the decreasing plasma impurities, which increased the breakdown voltage of the hydrogen gas. Using a compromise tune and increasing the inductance reduced the problem, but did not eliminate it [47].

Measuring the capacitive shift of the RF resonance with and without plasma allowed the estimation of the plasma inductance at ~0.15 μH, a significant fraction of the coil inductance of ~0.5 μH, but only ~4% of the total inductance. The problem was finally mitigated by starting each pulse with 1.96 MHz for ~5 μs before switching to 2 MHz, where the source is matched for maximum H⁻ beam current [48]. This reduced the ~30 μs H⁻ beam rise, seen in Fig. 9(a), to ~10 μs, seen in Fig. 9(b).

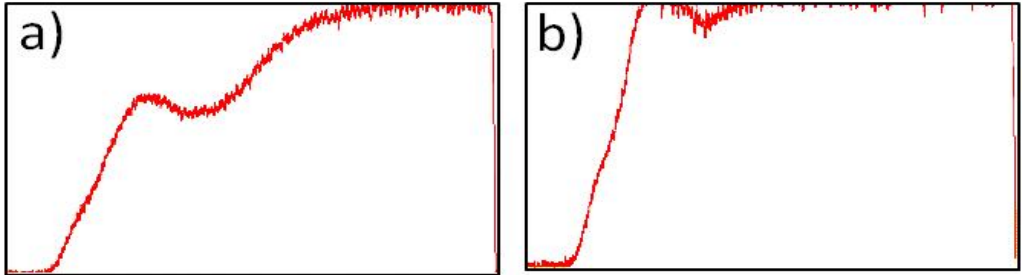

**Fig. 9:** The H⁻ beam rise time (a) for 2 MHz and (b) for a 1.96 MHz start followed by 2 MHz

Developing a 70 kV, 2 MHz transformer [49] allowed the 2 MHz amplifier to be moved from the 65 kV high-voltage platform enclosure to ground, which significantly reduced the RF amplifier problems [47].

The source is kept at a potential of −65 kV, which generates the extraction field. An 8 kV supply located on the −65 kV platform generates the −6.2 kV required to intercept most of the co-extracted electrons [46], which are driven to the side by the dipole dumping magnet.

There was also a significant learning curve for the LEBT [45], but its availability is very high since the summer of 2010. While models predict no significant losses, recently implemented thermocouples in the centre ground of the LEBT show significant heat generated by the beam as well as by occasional discharges.

## 8 RF antennas for high-current, high-duty-factor negative ion sources

High H⁻ beam currents require high radio-frequency (RF) power, with several hundreds of amperes flowing through the antenna. Hundreds of amperes induce oscillating voltages of the order of a kilovolt, which can cause sputtering of the antenna surface and large fluctuations in the plasma potential. Increasing the RF power increases the plasma density. This can cause an increasing fraction of the current to pass through the plasma, bypassing the ~0.5 μH antenna and leading to a saturation of the plasma density. Electrical insulation and water cooling are required for the high plasma densities needed to produce large H⁻ beam currents. This was initially achieved with a glass/enamel-coated Cu antenna [19].

In 1994 DESY acquired a copy of the LBNL RF source and spare antennas to test its suitability as an injector for the Hadron–Electron Ring Accelerator (HERA). All antennas were coated by the Porcelain Patch and Glaze Corporation (P&G) with a single layer of porcelain [50]. Unfortunately, each antenna developed one or several holes in apparently random locations, compromising the H⁻ beam output. Glassy deposits needed to be cleaned from the source before operation could restart with a new antenna. Of the 21 antennas, four failed during the first day of operation and another six failed over the next 14 days. One antenna survived for 167 days, as seen in Fig. 10(a) (from [22]). Converted to daily failure probabilities, after 19% failed on the first day, the probability dropped to ~3.3% for the next 20 days, and then to ~1.2% for the next 146 days. Most remarkable is the lack of old-age failures. This suggests that the sputtering of the bulk antenna coating was not the root cause of the failures.

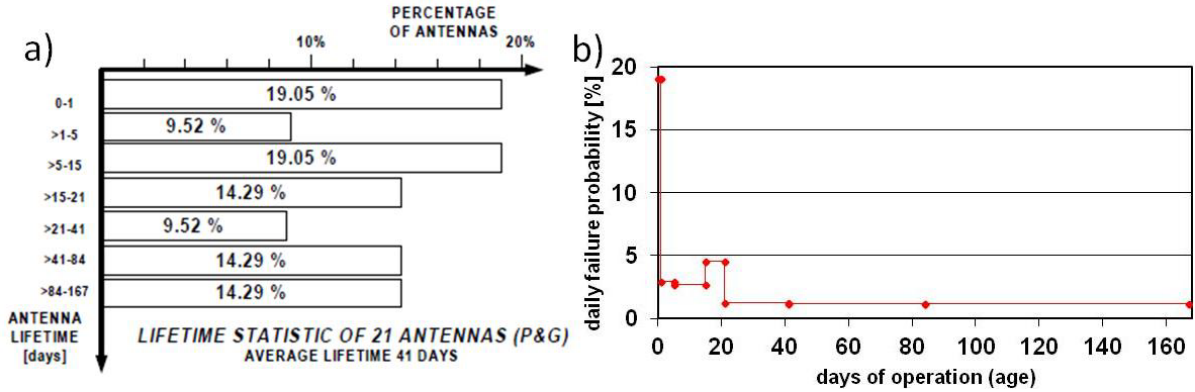

**Fig. 10**: Antenna lifetimes (a) as reported by DESY and (b) plotted as daily failure probability ((a) from [22])

When the construction of SNS started in 1999, multicusp ion sources with internal antennas were in use or being tested at several laboratories throughout the world. Reports of antenna failures raised concern over whether the H⁻ source developed for SNS could meet the SNS lifetime requirement. Naturally, each laboratory tested the antennas with their specific requirements, which raised the question of how such lifetimes should be scaled to the SNS requirements.

- If the limiting factor was the heating of certain parts, the average RF power would be the appropriate scaling factor. This meant that the 50 h observed at the Paul Scherrer Institute (PSI), Switzerland, with continuous 6–8 kW, 2 MHz discharges disqualified the P&G antennas [51]. However, the 250 h achieved at PSI [51] with a quartz antenna [52] was promising. The lifetime of more than 500 h of a Ti tube inside a quartz tube driven by a continuous 13 MHz, 2 kW discharge [53] could meet the SNS requirements if the required 50 mA H⁻ current could be produced with 30 kW or less.
- However, if the limiting factor was related to turn-on issues of the high-power RF, the lifetime had to be scaled with the repetition rate. Only DESY tested the antennas with pulsed high-power RF, and if 45 kW could produce the required 50 mA, scaling with the repetition rate reduces the observed 41-day average lifetime to 0.7 days at 60 Hz.
- Finally, if the limiting factor was related to the high-power RF, the lifetime should be scaled with the duty factor, which would reduce the observed DESY lifetime to 0.2 days at 60 Hz and 1 ms.

The DESY results clearly questioned the chances of the P&G antennas meeting the SNS 1 MW requirements. It was found that the single layer of porcelain was only ~0.1–0.2 mm thick, which can only withstand 1–2 kV. This is marginal compared to the voltages generated in the coil at high power. A microscopic cross-section revealed highly variable porosity, which is common in air-sprayed, air-fired porcelain. A chemical analysis showed a significant amount of $TiO_2$, which is a dielectric and reduces the voltage drop inside the porcelain [54].

The antenna lifetimes were extended by Cherokee Porcelain [55], coating Cu antennas with four or five layers of $TiO_2$-free porcelain, which yielded a ~0.7 mm thick, low-dielectric insulation. Operational experience showed practically no failures at low power and low duty factor. Antenna failures started above 3% duty factor and yielded approximately one failure in about seven source service cycles of each ~20-week run. After the duty factor during conditioning was increased to 7%, three antennas failed within hours, after which the conditioning duty factor was reduced to the production duty factor. When the production duty factor was raised to 5.4% and the RF power was raised to ~60 kW, the antenna failures increased to two or three failures in about seven or eight source service cycles of each ~20-week run [56].

One antenna failed after 22 days; all other failures occurred in the first 11 days, and half of the failures occurred in the first six days, despite increasing the source service cycle up to six weeks [56]. This is consistent with infant mortality and the absence of old-age failure. This is also consistent with plasma emission spectra showing that Na sputtered from the antenna only during the first day when the plasma contains heavy impurities, initially water and later Cs [2]. It is also consistent with the removed antennas being blackened by carbon, most likely emitted from the stainless heat shield. The carbon becomes a sputter-resistant antenna coating after the impurities disappear from the plasma.

A more recent analysis revealed that most failures occur at the end of the coil where the antenna tube is exposed to the plasma. An improved geometry is being developed [48]. More recently, antenna failures have been significantly reduced by excluding all antennas with visible or tangible surface imperfection from production runs. In addition, efforts are under way to improve the cleanliness of the coating process [57].

Why not simply adopt DESY's external antenna scheme? The reason is because SNS's 67 times larger duty factor puts a significant heat load on the plasma chamber. SNS started to develop $H^-$ sources with external antennas in 2003 [58]. Several designs, some of which included a Faraday shield, did not bring the anticipated results. After two $Al_2O_3$ chambers failed, as expected from model calculations, the effort adopted aluminium nitride (AlN) because of its much higher thermal conductivity. Early in 2009 the AlN $H^-$ source was approved for neutron production. However, within about five weeks, five AlN source failures forced source changes, and the internal antenna source had to be reinstated to restore acceptable availability. Another serious issue is continuous beam loss during the entire one- to two-week service cycles, which is now suspected to be caused by emissions from AlN [57].

## 9  Caesium and the management of $Cs_2CrO_4$ cartridges

Its 260 pm atomic radius makes caesium the largest naturally occurring atom, as seen in Fig. 11 [59]. When adsorbed on the surface of another metal, its outermost electrons mix with the conduction electrons, forming strong, ionic-like bonds with the surface. Accordingly, as an adsorbate, the ionic radius of 181 pm becomes more relevant and is shown as a dashed line in Fig. 11. The line shows that there is a significant mismatch between the Cs ionic radius and the atomic radius of metals typically used in ion sources.

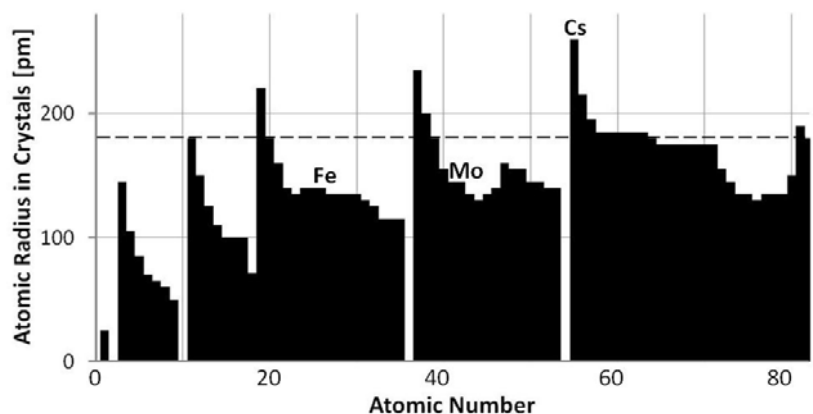

**Fig. 11:** Atomic radii versus atomic number. The dashed line shows the ionic radius of Cs

At low doses Cs adsorbs at the most attractive locations of the metal substrate, but the radius mismatch forces additional Cs to adsorb at less and less attractive locations. This causes the surface bond energy to decrease with increasing Cs coverage. This can be seen in Fig. 12, which shows functions fitted to data for Cs on a polycrystalline Mo surface [60] and on a 110 Mo surface [61].

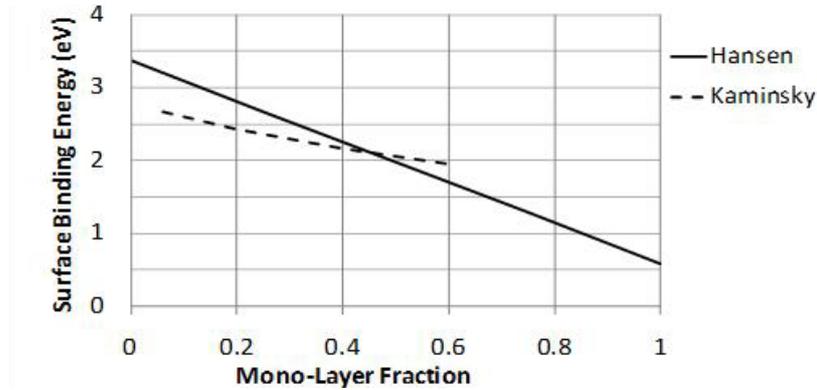

**Fig. 12**: Surface binding energy for Cs on polycrystalline Mo (Kaminsky [60]) and 110 Mo (Hansen [61]) versus the Cs coverage.

Cs atoms that adsorb on top of a Cs layer have small binding energies, and therefore are very rapidly emitted at room temperature. The decreasing binding energy of the first monolayer has an interesting effect on its thermal emission. For constant bond energies, thermal emission emits a certain fraction of the remaining adsorbates, causing an exponential decay of the adsorbate population. However, because the bond energy increases with a decreasing population, the bond energy practically stabilizes the population at a certain fraction, which depends on the temperature. This is not quite obvious from Fig. 13, which shows the expected fractions for different temperatures versus the time on a logarithmic scale calculated with the Hansen approximation [62].

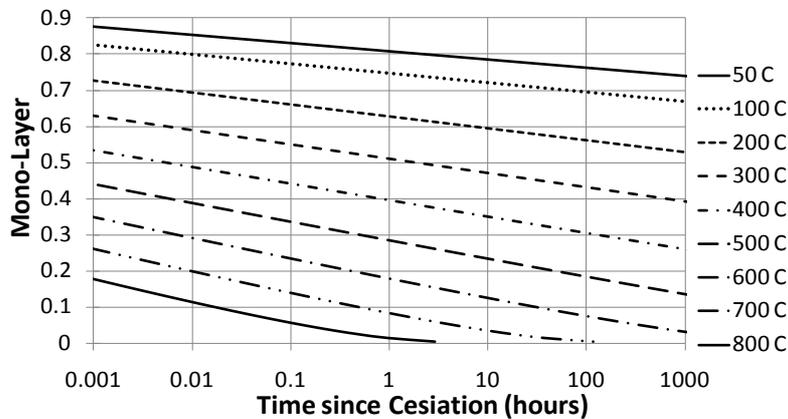

**Fig. 13:** The Cs surface coverage versus time for different surface temperatures

However, using linear time scales, Fig. 14 [56] shows that significant changes only happen at the beginning and then become too small to be noticed: looking at a scale of 5 h in Fig. 14(a), the significant change happens in the first half hour, and looking on a scale of 35 days in Fig. 14(b), the significant change occurs in the first few days. This means that one can use a constant temperature to approximately control the Cs population on a metal surface. This explains why successful caesium-enhanced volume H⁻ sources control the temperature of the surfaces near the outlet. The temperature is adjusted to obtain the fractional monolayer that yields the most H⁻ beam.

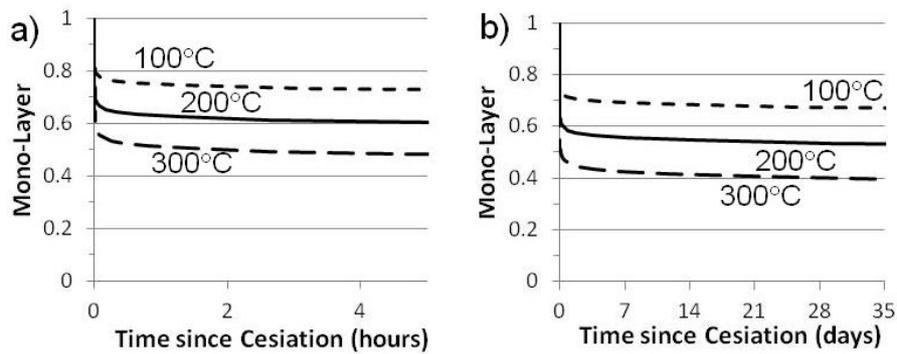

**Fig. 14:** Same as Fig. 13, but linear time scales of (a) 5 h and (b) 35 days

Cs has an ionization energy of 3.9 eV and an electron affinity of 0.47 eV, and therefore remains mostly neutral in the cold plasma near the outlet. Accordingly, some of the Cs escapes the source as neutral atoms and adsorbs on surfaces that are in line of sight with the outlet. For the SNS H⁻ source, these surfaces include the extractor, the first lens and the ground electrode. The adsorbed Cs lowers the work function, which leads to electron emission and discharges from the negatively charged first lens, which can make the LEBT temporarily inoperable. Lowering the use of Cs has significantly improved the performance of the first lens [62].

Figure 15(a) shows the cross-section of a $Cs_2CrO_4$ cartridge [63]. Eight such cartridges, containing slightly less than 30 mg of Cs, are inserted into the Cs collar, shown in Fig. 15(b). The figure also shows the water-cooled filter magnets, the 7 mm diameter source outlet, and the Mo converter, which was introduced in 2007 to boost the H⁻ current output [45].

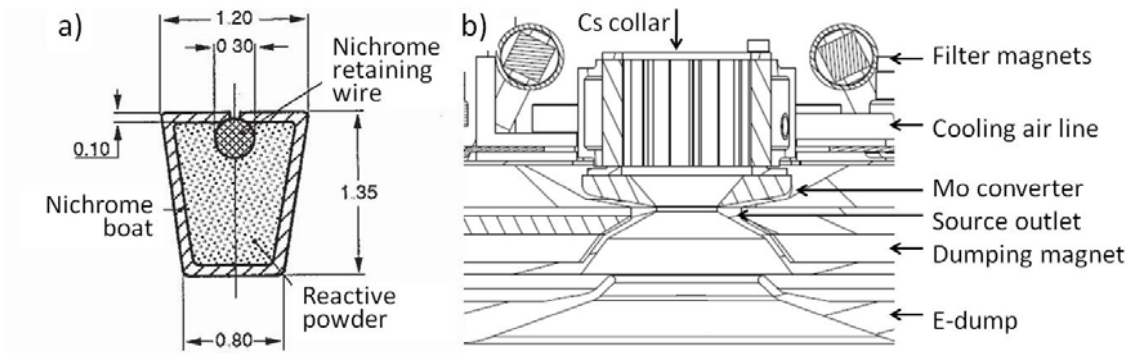

**Fig. 15:** (a) $Cs_2CrO_4$ cartridge cross-section and (b) the cartridge holding Cs collar with the SNS converter

The mixture of $Cs_2CrO_4$ and St101 (a getter made of 16% Al and 84% Zr) is a reactive, fine powder having a very large surface area covered with adsorbates. SNS found that the cartridges can be fully degassed without emitting large amounts of Cs by heating them for 3 h to ~250°C. This is confirmed by the residual gas mass analyser (RGA), which shows the same partial pressures before and after caesiation.

Without sufficient degassing, the St101 will getter the surface sorbates before starting to reduce the $Cs_2CrO_4$, and so release less, and sometimes insufficient, Cs. Insufficient degassing is confirmed by the lower partial pressures of masses 18, 28 and 44 after caesiation, because the caesiation stopped the outgassing from the cartridges.

For Cs to form strong, ionic bonds with the converter, the Mo converter must also be cleaned of sorbates. SNS found that 3 h of 50 kW plasma at a 5.3% duty factor sputter-cleans the Mo converter, and accordingly it coincides with the conditioning of the cartridges and the high-voltage conditioning of the two electrostatic lenses.

In summary, the SNS source is installed and evacuated while being leak-checked. When complete, the protective cages are closed, the collar is heated to ~250°C, and the 13 MHz plasma is started, followed by the 2 MHz. After 3 h of conditioning and ramping up the lens voltages, the collar temperature is raised to 550°C. After 12 min the temperature is lowered to ~180°C to retain a near-optimal fractional Cs layer. The extracted H⁻ beam typically grows for the first few days as the Cs fraction decreases from above optimal towards the optimal value.

## 10  Persistent H⁻ beams and the plasma potential

Despite short pulses, low repetition rates and a cold air-cooled Cs collar, the H⁻ beam decayed in the first neutron production runs, with the first run shown in Fig. 16. However, the beam could be restored by raising the temperature of the Cs collar for a certain time period.

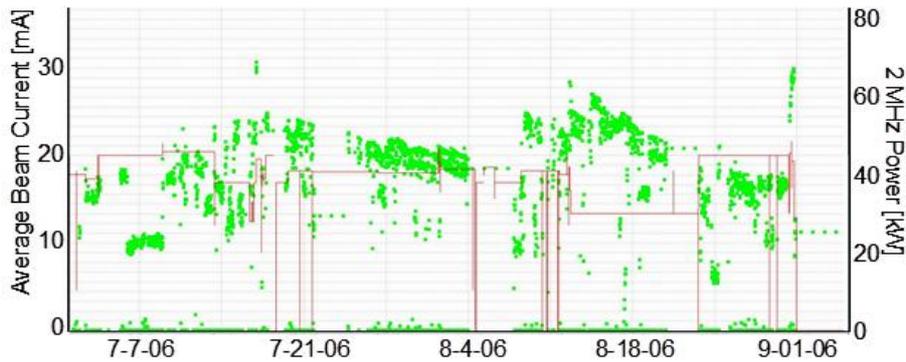

**Fig. 16:** Decaying H⁻ beam during the first neutron production run

Caesiations yielded unpredictable results until the need for degassing the cartridges was recognized in 2007. Even then, the H⁻ beam decayed during the first night, but became persistent after recaesiating the following morning [56]. Only in 2008, after the plasma conditioning was extended to 3 h, did the H⁻ beams become persistent after the first caesiation. In 2009, to meet the growing H⁻ current requirement, the Cs collar temperature was raised without noticing beam loss during the three-week and later four-week source service cycles. This lead to the calculations presented in the last section [62]. After a contamination event in 2011, the source service cycle of the uncontaminated source was extended to six weeks without observing decay in the H⁻ beam [2].

However, a thermal air leak during the summer of 2011 caused the H⁻ beam to decay by ~1% per hour. The beam was restored with recaesiations, after which it decayed again with the same rate [56]. This is consistent with the sputtering of the Cs by the heavy air ions and its subsequent loss from the plasma. Fig. 17 shows the ratio of the Cs sputter threshold to the Cs surface bond energy versus the square root of the mass of the sputtering ion, using the Bohdansky approximation [64]. The figure indicates that water and CO ions have the most effective masses for sputtering Cs, while $N^+$ and $NH_3^+$ need just a little more energy to sputter Cs. On the other hand, hydrogen ions are much too light to sputter Cs effectively.

Assuming surface bond energies of 2 eV for Cs on Mo [60], the fact that pure hydrogen plasma fails to sputter Cs suggests that the plasma potential is less than 25 V, whereas the fact that nitrogen ions sputter Cs suggests that the plasma potential exceeds 8 eV.

RGA data from the leaky source suggested roughly a $10^{-6}$ cm³ s⁻¹ air leak. This is consistent with a plasma impurity of ~1 ppm, which would suggest that the SNS hydrogen plasma has normally less than $10^{-8}$ impurities when it provides persistent beams for six weeks [2]. At this time it is unclear how that can be reconciled with the use of ultra-high-purity hydrogen (<10 ppm impurities) without further purification.

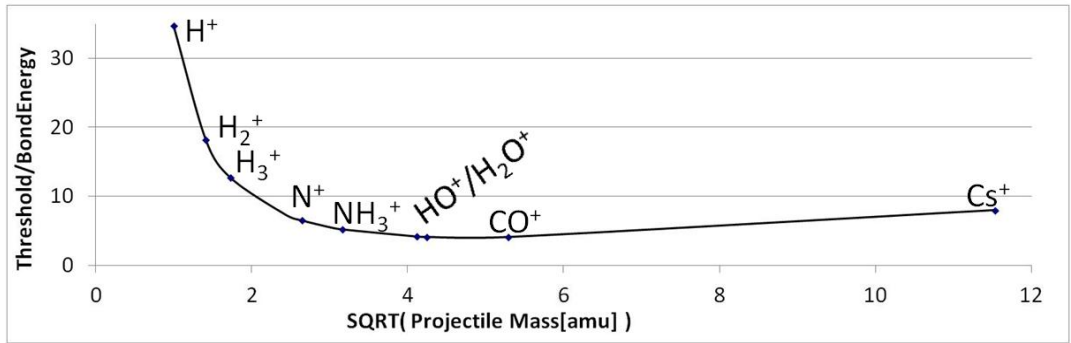
**Fig. 17:** Cs sputter threshold to surface bond energy ratio for various ions

## 11   The caesium consumption

Many papers have pointed out that H is too light for sputtering Cs. This is not exactly true because sputtering depends on the energy of the hydrogen particles. On the other hand, it is well known that Cs-enhanced H¯ sources require a small flux of Cs to maintain a constant output of the H¯ beam. SNS appears to have the first caesiated H¯ source that is able to run at high duty factor for long time periods without a continuous supply of Cs [46].

The Cs consumption is difficult to determine because it depends on the design and even more on the operation of the Cs system. Starting up a new system, Cs should be generously dispensed to be sure to reach the optimal Cs distribution within a reasonable time period, normally judged by the H¯ beam current output. However, once the H¯ output peaks, the Cs dispense rate should be drastically reduced for sustainable operations. The details depend on the temperatures of certain surfaces. There is a long learning curve of adjustments to maintain that maximum H¯ output current by replacing the Cs that is lost and not more. After sustainability is established, the learning curve begins on how to reach the optimal caesiation in a minimum of time with a minimum of Cs.

An excellent example is the HERA magnetron source, which used 6 mg of Cs when it was initially started in 1983. In its last run of 2008, that amount was reduced to 0.7 mg per day [65].

Another example is the multicusp, filament-driven LANSCE H¯ source, where the outlet-facing Mo converter features −300 V to boost the extraction of the converter-surface-produced H¯ ions. This voltage is sufficient to empower hydrogen ions to sputter Cs, which could contribute to the consumption of 0.7 g of Cs per day operating at a 5% duty factor [66]. This exceeds the HERA consumption 1000-fold. As discussed, thermal emission initially contributes to Cs losses, but if properly adjusted, those losses are marginalized by self-stabilization within a few days. Eventually Cs losses are dominated by sputter losses, and therefore only occur when plasma is present. Accordingly, the Cs consumption should be normalized to the plasma duty factor (5%), which yields for the LANSCE source ~14 g per plasma-day (daily consumption divided by the plasma duty factor), where a plasma-day is the length of time needed to accumulate a full day of plasma operation.

Normalizing the HERA source with its 0.0075 % duty factor yields ~10 g per plasma-day [65], not very different from the LANSCE source.

After heating the SNS cartridges three times for 30 min to 550°C, an analysis showed the Cs to be fully depleted. That suggests that ~4 mg Cs is released during the first 12 min. Producing H¯ beam for up to six weeks suggests a daily consumption of <0.1 mg per day. Normalizing with the 5.3% duty factor yields <1.8 mg per plasma-day, almost four orders of magnitude less than the other sources.

## 12 The co-extracted electrons, a real challenge for high-power H⁻ sources

Negative ion sources are placed on negative high-voltage platforms, which generate the field to accelerate the escaping negative ions towards the near-ground extractor. Naturally this field also accelerates all electrons that venture to the meniscus, forming an electron beam. While this is a small inconvenience for low-current and/or low-duty-factor H⁻ sources [22], it becomes a challenging problem at high power and high duty factor [67] because there are normally many more electrons than H⁻. Co-extracted electron currents exceeding the H⁻ current by a factor between 100 and 200 have been reported [18], although the introduction of the outlet collar and the use of NdFe filter magnets reduced this ratio to ~50 [19]. Further optimizations can lower this value by another factor of 2 [21, 68, 69].

While Cs normally increases the H⁻ beam current, it also decreases the electron current by a factor of two [40], which brings the electron to H⁻ ratio typically into the single-digit range [68]. When extracting ~50 mA H⁻ beam, the SNS source co-delivers only about 20 mA of electrons, which is surprisingly small.

Accelerating amps of electron currents from uncaesiated sources or fractions of amps from caesiated H⁻ sources would generate destructive electron beams with kilowatts of power, while posing a destabilizing excessive burden on the extraction supply. The common solution is to stop the electron beam at an intermediate electrode, powered by a high-current, reduced-voltage supply located on the high-voltage platform. This decouples the energy of the H⁻ beam from fluctuations of the electron beam, which can be a serious problem for pulsed H⁻ sources.

Separating the electron beam from the H⁻ beam requires a magnetic field, which will also deflect the H⁻ beam by a small angle of 2 to 3°. In the SNS source, this angle is compensated by tilting the source by a few degrees. As shown in Figs. 4(b) and 6(b), other sources typically use two opposing dipole fields to replace the angular deflection with small doglegs in the ion trajectories.

For smaller source voltages, the e-dump can be placed beyond the grounded extractor [26], although energy-reducing voltages will focus the ion beam. The SNS source (see Fig. 8(a)), the J-PARC source (see Fig. 6(b)) and the TRIUMF source (see Fig. 4(b)) all place the e-dump between the source and the extractor, which minimizes focusing. However, this couples the energy of the electron beam and the extraction field near the outlet, which is not ideal.

For high-current, high-duty-factor H⁻ sources, it is essential to properly model a robust dumping of the electron beam, which will determine the voltage of the e-dump. Placing an extraction or 'puller' electrode between the source outlet and the e-dump allows the control of the extraction field while dumping the electrons at a safe voltage [70].

## Acknowledgements

This manuscript was prepared at Oak Ridge National Laboratory, which is managed by UT-Battelle, LLC, under contract DE-AC05-00OR22725 for the US Department of Energy. The author appreciates the invaluable comments from M. Bacal, J. Galambos, J. Holmes and H. Oguri.